\documentclass[twocolumn,superscriptaddress,aps,preprintnumbers,
amsmath,amssymb,prl,nofootinbib,showpacs]{revtex4}

\usepackage{epsfig}
\usepackage{amsmath}
\usepackage{amssymb}
\usepackage{subfigure}
\usepackage{cancel}
\usepackage{textcomp}
\usepackage{calc}
\usepackage{graphicx}
\usepackage{dcolumn}
\usepackage{color}

\begin{document}

%%%%%%%%%%%%%%%%%%%%%%%%%%%%%%%%%%%%%%%%%%%%%%%%%%%%%%%%%%%%%%%%%%%%%%%%%%%%%%%%%%%%%%%%%%%%%%%%%%%%

\title{Leptogenesis via Axion Oscillations after Inflation}

\author{Alexander Kusenko}
\affiliation{Department of Physics and Astronomy, University of California, Los
Angeles, California 90095-1547, USA}
\affiliation{Kavli IPMU (WPI), UTIAS, The University of Tokyo, Kashiwa, Chiba 277-8583, Japan}
\author{Kai Schmitz}
\email[Corresponding author. E-mail: \vspace{0.2cm}]{kai.schmitz@ipmu.jp}
\affiliation{Kavli IPMU (WPI), UTIAS, The University of Tokyo, Kashiwa, Chiba 277-8583, Japan}
\author{Tsutomu T.~Yanagida}
\affiliation{Kavli IPMU (WPI), UTIAS, The University of Tokyo, Kashiwa, Chiba 277-8583, Japan}

%%%%%%%%%%%%%%%%%%%%%%%%%%%%%%%%%%%%%%%%%%%%%%%%%%%%%%%%%%%%%%%%%%%%%%%%%%%%%%%%%%%%%%%%%%%%%%%%%%%%

\begin{abstract}
Once a light axionlike scalar field couples to the electroweak gauge bosons, its classical
motion during reheating induces an effective chemical potential for the fermion number.
In the presence of rapid lepton number ($L$)-violating processes in the plasma,
such a chemical potential provides a favorable opportunity for baryogenesis
via leptogenesis.
We are able to demonstrate that $L$ violation due to the exchange of heavy
Majorana neutrinos is sufficient for a successful realization of this idea.
Our mechanism represents a novel and minimal alternative to thermal leptogenesis,
which turns out to be insensitive to the masses and $CP$-violating phases
in the heavy neutrino sector.
It is consistent with heavy neutrino masses close to the scale of grand
unification and, quite complementary to thermal leptogenesis, requires
the reheating temperature to be at least of order~$10^{12}\,\text{GeV}$.
\end{abstract}

%%%%%%%%%%%%%%%%%%%%%%%%%%%%%%%%%%%%%%%%%%%%%%%%%%%%%%%%%%%%%%%%%%%%%%%%%%%%%%%%%%%%%%%%%%%%%%%%%%%%

\date[Received ]{18 December 2014; revised manuscript received 18 February 2015; published 2 July 2015}
\pacs{98.80.Cq, 11.30.Fs, 14.60.St, 14.80.Va}
\maketitle
\preprint{IPMU 14-0352}

%%%%%%%%%%%%%%%%%%%%%%%%%%%%%%%%%%%%%%%%%%%%%%%%%%%%%%%%%%%%%%%%%%%%%%%%%%%%%%%%%%%%%%%%%%%%%%%%%%%%

The Peccei-Quinn (PQ) solution to the strong $CP$ problem~\cite{Peccei:1977hh} has
led to the prediction of a light scalar field called the axion~\cite{Weinberg:1977ma},
which appears generically in a number of models~\cite{Kim:1979if,Kim:1986ax} and which
has a characteristic coupling to the gauge fields of the form $f_a^{-1}a \,F\tilde{F}$,
where $f_a$ is the scale of PQ symmetry breaking.
However, the motivation for considering axionic fields extends well beyond
the context of the strong $CP$ problem.
Axions are ubiquitous in string theory, where at least one such field is
generically associated with the Green-Schwarz mechanism of anomaly
cancellation~\cite{Green:1984sg} and the scale $f_a$ of which lies a few
orders of magnitude below the Planck scale~\cite{Svrcek:2006yi}; but multiple
other axions can also appear.
While the model-independent axion is coupled to all gauge
groups with a universal coupling strength, the additional axionic fields
can couple to different groups with couplings that depend on both the gauge
group and the particle content of the model~\cite{Ibanez:1986xy}.
The masses of these model-dependent axions can be different as they arise
from their couplings to different anomalous groups.
We will focus, in particular, on the axion (or linear
combination of axions) that has a coupling to the
electroweak $SU(2)$ gauge fields.

%%%%%%%%%%%%%%%%%%%%%%%%%%%%%%%%%%%%%%%%%%%%%%%%%%%%%%%%%%%%%%%%%%%%%%%%%%%%%%%%%%%%%%%%%%%%%%%%%%%%

During inflation, light scalar fields develop large expectation values~\cite{Bunch:1978yq}.
The relaxation of the axion field to the minimum of its effective potential begins once
the Hubble rate becomes comparable to the axion mass. 
While the field $a(t)$ is slowly rolling towards its origin, its coupling to
the $SU(2)$ gauge fields and, via the anomaly, to the fermionic current
$j^\mu =\bar{\psi}\gamma^\mu\psi$ induces an effective $CPT$-violating term
$a(t) F\tilde{F} \propto (\partial_t a(t))\, j^{0} $, which serves as a
chemical potential for fields carrying nonzero baryon or lepton number.%
\footnote{This illustrate that, while the axion could equally couple to
the hyercharge gauge boson, a coupling to the standard model gluons would, by contrast,
not provide a sufficient basis for leptogenesis.}
Then, in the presence of rapid lepton number-violating processes in the plasma---for example,
due to the exchange of virtual heavy right-handed neutrinos---the conditions
for successful leptogenesis are satisfied.
Here, one important detail is that lepton number violation needs to occur at a fast rate,
$\Gamma_L \gg \dot{a}/a$, so that the axion field acts
as an adiabatic background during leptogenesis.
It is therefore essential that the generation of the lepton asymmetry be driven
by an external source, i.e., by scatterings with heavy neutrinos in the intermediate
state in our case.
If the rate of lepton number violation was instead tied to the change
in the axion field value, we would, by contrast, not be able
to interpret the axion velocity as an effective chemical potential~\cite{Dolgov:1994zq}.
As we will see, a consequence of the requirement of a fast rate $\Gamma_L$ is, in particular,
that we will have to adopt large values for the reheating temperature,
$T_{\rm rh} \gtrsim 10^{12}\,\textrm{GeV}$.

%%%%%%%%%%%%%%%%%%%%%%%%%%%%%%%%%%%%%%%%%%%%%%%%%%%%%%%%%%%%%%%%%%%%%%%%%%%%%%%%%%%%%%%%%%%%%%%%%%%%

A similar scenario was discussed in connection with a flat direction that
carries no baryon or lepton number~\cite{Chiba:2003vp}.
Our treatment of the asymmetry is different, and we obtain very different results.
Our scenario is also similar to leptogenesis via Higgs relaxation~\cite{Kusenko:2014lra},
where the Higgs coupling to $F\tilde{F}$ is assumed to arise from a higher-dimensional
operator, unlike in the present scenario, where the required coupling appears generically
for any axion coupled to the electroweak gauge fields.
One can also draw an analogy to models of spontaneous baryogenesis~\cite{Cohen:1987vi},
in particular to realizations of spontaneous baryogenesis during the
electroweak phase transition (EWPT)~\cite{Dine:1990fj}.
Here, the Higgs field in the expanding bubble wall generates an effective chemical potential
for the fermions.
Our scenario is different in that the ``wall'' is represented by an
axionic field moving in the timelike direction uniformly in space, unlike the 
bubble wall moving in a spacelike direction.
Finally, we note that, in the context of electroweak baryogenesis, also the
QCD axion may be employed to generate an effective chemical potential~\cite{Kuzmin:1992up}.
Successful baryogenesis then requires that
the EWPT be delayed below the GeV scale; for instance,
due to some additional Higgs-dilaton coupling~\cite{Servant:2014bla}.

%%%%%%%%%%%%%%%%%%%%%%%%%%%%%%%%%%%%%%%%%%%%%%%%%%%%%%%%%%%%%%%%%%%%%%%%%%%%%%%%%%%%%%%%%%%%%%%%%%%%

Provided that the axion $a$ is to be identified with the pseudo-Nambu-Goldstone boson of a
spontaneously broken $U(1)$ symmetry with a compact global topology, its initial value
at the end of inflation is $a_0=f_a \theta_0$, where the angle $\theta_0$ takes a
random value in the range $\theta_0 \in [0,2\pi)$.
Assuming that the PQ symmetry is broken sufficiently early before the end of inflation
(and is not restored during reheating), this initial value 
ends up being constant on superhorizon scales.
For definiteness, we set $a_0 = f_a$ and treat $f_a$
as a free parameter in the following.
Anticipating that the final baryon asymmetry will depend on $a_0$, we require that the
baryonic isocurvature perturbations induced by the quantum fluctuations of the axion field
during inflation be smaller than the observational upper limit.
This implies a constraint on the Hubble rate during inflation: 
$H_{\rm inf}/(2\pi)/a_0 \lesssim 10^{-5}$~\cite{isocurvature}, or 
\begin{align}
H_{\rm inf} \lesssim 6 \times 10^{11}\,\text{GeV}
\left(\frac{f_a}{10^{15}\,\text{GeV}}\right) \,.
\label{eq:Hbound}
\end{align}

%%%%%%%%%%%%%%%%%%%%%%%%%%%%%%%%%%%%%%%%%%%%%%%%%%%%%%%%%%%%%%%%%%%%%%%%%%%%%%%%%%%%%%%%%%%%%%%%%%%%

The evolution of the homogeneous axion field in its effective potential
$V_{\rm eff}$ around the origin is described by
\begin{align}
\ddot{a} + 3\, H \dot{a} = - \partial_a V_{\rm eff} \,, \quad
V_{\rm eff} \approx \frac{1}{2} m_a^2 a^2 \,,
\label{eq:axionEOM}
\end{align}
where we have neglected the backreaction of lepton number generation on the evolution
of the axion field.
$m_a$ denotes the axion mass, which we assume to arise via dimensional transmutation,
i.e., from an additional coupling of the axion to the gauge fields
of some strongly coupled hidden sector.
Given a dynamical scale $\Lambda_H$ in this hidden sector, the axion mass is then
of $\mathcal{O}\left(\Lambda_H^2 / f_a\right)$.
For consistency, we require $m_a$ to be smaller than $H_{\rm inf}$,
the Hubble rate at the end of inflation: 
\begin{align}
m_a \lesssim H_{\rm \inf} \,.
\label{eq:mabound}
\end{align}
When inflation is over, the axion field remains practically at rest until
the Hubble parameter drops to $H_{\rm osc}=m_a$.
Once the axion field is in motion, the effective Lagrangian contains the term 
\begin{align}
\mathcal{L}_{\rm eff} &
\supset \frac{g_2^2}{32\pi^2}\frac{a(t)}{f_a} F\tilde{F} 
= - \frac{a(t)}{N_f f_a} \partial_\mu \left(\bar{\psi}\gamma^\mu\psi \right)
\label{eq:mueff1}\\
& = \frac{\partial_t a(t)}{N_f f_a} \left (\bar{\psi}\gamma^0\psi \right) + \cdots
=  \mu_{\rm eff}\, j^0 + \cdots \,,
\label{eq:mueff} 
\end{align}
with $g_2$ being the $SU(2)$ gauge coupling and $N_f = 3$ the
number of fermion generations in the standard model, where we have
used the anomaly equation in Eq.~\eqref{eq:mueff1}, and integration
by parts in Eq.~\eqref{eq:mueff}.
In the following, we will absorb $N_f$ in our definition of $f_a$
and simply determine the effective chemical
potential as $\mu_{\rm eff} = \dot{a} / f_a$.

%%%%%%%%%%%%%%%%%%%%%%%%%%%%%%%%%%%%%%%%%%%%%%%%%%%%%%%%%%%%%%%%%%%%%%%%%%%%%%%%%%%%%%%%%%%%%%%%%%%%

Now the necessary conditions for generating a lepton asymmetry are satisfied.
A nonzero effective chemical potential shifts the energy levels of
particles as compared to antiparticles.
If lepton number is not conserved, the minimum of the free energy
in the plasma is reached for a different number density of leptons than for
antileptons, i.e., for $n_L \equiv n_{\ell} - n_{\bar{\ell}} \neq 0$.
Instead, if the lepton number violation is very rapid, the minimum of the
free energy is obtained for an equilibrium number density of%
\footnote{In the following, we shall approximate all number and energy
densities by their corresponding expressions in the classical
Boltzmann approximation.
We will only take care of quantum-statistical effects when counting
relativistic degrees of freedom.}
\begin{equation}
n_L^{\rm eq} = \frac{4}{\pi^2} \, \mu_{\rm eff}\, T^2.
\end{equation}

%%%%%%%%%%%%%%%%%%%%%%%%%%%%%%%%%%%%%%%%%%%%%%%%%%%%%%%%%%%%%%%%%%%%%%%%%%%%%%%%%%%%%%%%%%%%%%%%%%%%

Lepton number violation is mediated by the exchange of heavy neutrinos.
In contrast to thermal leptogenesis~\cite{Fukugita:1986hr}, we will assume
all heavy right-handed neutrino masses to be close to the scale of grand unification (GUT),
$M_i \sim \mathcal{O}\left(10^{-1} \cdots 1\right) \Lambda_{\rm GUT}
\sim 10^{15} \cdots 10^{16}\,\text{GeV}$, so that the heavy neutrinos
are never produced thermally, i.e., $T \ll M_i$ at all times.
This assumption serves the purpose to separate the mechanism under study from
the contributions from ordinary thermal leptogenesis.
In the expanding universe, the evolution of the $L$ number density $n_L$
is then described by the Boltzmann equation 
\begin{align}
\dot{n}_L + 3\, H n_L \simeq - \Gamma_L
\left(n_L - n_L^{\rm eq}\right) \,,\quad
\Gamma_L = 4 \,n_{\ell}^{\rm eq} \,\sigma_{\rm eff} \,,
\label{eq:LBoltz}
\end{align}
where $n_{\ell}^{\rm eq} = 2/\pi^2\,T^3$ and with
$\sigma_{\rm eff} \equiv \left<\sigma_{\Delta L = 2}\, v\right>$ denoting
the thermally averaged cross section of
two-to-two scattering processes with heavy neutrinos
in the intermediate state that violate lepton number by two units,
\begin{align}
& \Delta L = 2 : \quad \ell_i\ell_j \leftrightarrow HH \,, \:\:
\ell_i H \leftrightarrow \bar{\ell}_j\bar{H} \,, \\\nonumber
& \ell_i^T =
\begin{pmatrix}
\nu_i & e_i 
\end{pmatrix} \,,\:\: H^T =
\begin{pmatrix}
h_+ & h_0 
\end{pmatrix} \,, \:\: i,j = 1,2,3 \,.
\end{align}
We note that the term proportional to $n_L^{\rm eq}$
now acts as a novel production term for the lepton asymmetry,
as long as the axion field is in motion.  
For center-of-mass energies much smaller than the heavy neutrino
mass scale, $\sqrt{s} \ll M_i$, the effective cross section $\sigma_{\rm eff}$
is practically fixed by the experimental data on the light neutrino
sector~\cite{Buchmuller:2004nz}, assuming the seesaw mass matrix~\cite{seesaw}: 
\begin{align}
\sigma_{\rm eff} \approx
\frac{3}{32\pi} \frac{\bar{m}^2}{v_{\rm ew}^4} \simeq
1 \times 10^{-31} \,\text{GeV}^{-2} \,, \:\:
\bar{m}^2 = \sum_{i=1}^3 m_i^2 \,, 
\label{eq:sigmaeff}
\end{align}
where $v_{\rm ew} \simeq 174 \,\text{GeV}$ and where we have assumed that  
the sum of the light neutrino masses squared is of the same order
of magnitude as the atmospheric neutrino mass difference,
$\Delta m_{\rm atm}^2 \simeq 2.4 \times 10^{-3}
\,\textrm{eV}^2$~\cite{Agashe:2014kda}.

%%%%%%%%%%%%%%%%%%%%%%%%%%%%%%%%%%%%%%%%%%%%%%%%%%%%%%%%%%%%%%%%%%%%%%%%%%%%%%%%%%%%%%%%%%%%%%%%%%%%

For $a_0 \ll M_{\rm Pl}$, and as long as $H \gg m_a$, i.e., prior to the
onset of the axion oscillations, the axion energy density $\rho_a$ is much
smaller than the total energy density
$\rho_{\rm tot}=\rho_\varphi+\rho_R+\rho_a \approx \rho_\varphi+\rho_R$, where  
$\rho_\varphi$ and $\rho_R$ are the energy densities of the inflaton and of radiation.  
Reheating is described by a system of equations: 
\begin{align}
\dot{\rho}_\varphi + 3\,H \rho_\varphi  \: = -\, \Gamma_\varphi \rho_\varphi \,, & \quad 
\dot{\rho}_R + 4\,H \rho_R  = +\, \Gamma_\varphi \rho_\varphi \,, 
\label{eq:phiradBoltz}\\
\label{eq:Friedmann}
H^2   \equiv \big(\dot{R}/R\big)^2   = \frac{\rho_{\rm tot}}{3M_{\rm Pl}^2} 
\,,
& \quad \rho_{\rm tot}  \approx \left(\rho_\varphi + \rho_R \right) \, ,
\end{align}
where $\Gamma_\varphi$ is the inflaton decay rate.
The inflaton must not decay before the end of inflation, which implies 
\begin{align}
\Gamma_\varphi \lesssim H_{\rm inf} \,.
\label{eq:Gammabound}
\end{align}

%%%%%%%%%%%%%%%%%%%%%%%%%%%%%%%%%%%%%%%%%%%%%%%%%%%%%%%%%%%%%%%%%%%%%%%%%%%%%%%%%%%%%%%%%%%%%%%%%%%%

The rough temperature scale of leptogenesis as well as the axion mass scale in
our scenario  are determined by the requirement that the heavy neutrino-mediated $\Delta L = 2$
interactions must be in thermal equilibrium before the onset of axion oscillations,
$\Gamma_L \gg H \gtrsim m_a$, which yields
$T \sim T_L = g_*^{1/2}/\left(\pi\,\sigma_{\rm eff} M_{\rm Pl}\right) \sim 10^{13}\,\textrm{GeV}$
and $m_a \sim \sigma_{\rm eff} T_L^3 \sim 10^8\,\textrm{GeV}$.
Upon closer examination,
the solution for the temperature, $T^4 \equiv \pi^2/3/g_*\,\rho_R$,
according to Eqs.~\eqref{eq:phiradBoltz} and 
\eqref{eq:Friedmann} shows the following characteristic
behavior: within roughly one Hubble time after the end of inflation,
$T$ quickly rises to its maximal value,
\begin{align}
T_{\rm max} \simeq 5 \times 10^{13}\,\text{GeV}
\left(\frac{\Gamma_\varphi}{10^9\,\text{GeV}}\right)^{1/4}
\hspace{-3pt}\left(\frac{H_{\rm inf}}{10^{11}\,\text{GeV}}\right)^{1/2}
\hspace{-12pt},\hspace{-3pt}
\end{align}
after which the temperature decreases because the energy density
is dominated by the inflaton oscillations (which scale as matter).
During reheating, the temperature drops as $T \propto R^{-3/8}$
until radiation comes to dominate at time  $t=t_{\rm rh} \simeq \Gamma_\varphi^{-1}$, when
$\rho_R = \rho_\varphi$, and the reheating temperature is  
\begin{align}
T_{\rm rh} \simeq 2 \times 10^{13}\,\text{GeV}
\left(\frac{\Gamma_\varphi}{10^9\,\text{GeV}}\right)^{1/2} \,.
\label{eq:TRH}
\end{align}
After the end of reheating, i.e., for $t > t_{\rm rh}$,
the expansion is then driven by relativistic radiation and the
temperature simply decreases adiabatically, $T \propto R^{-1}$.
In the case of a large axion decay constant, this phase of radiation
domination, however, does not last all the way to the time of
primordial nucleosynthesis. 
Instead, the axion comes to dominate the total energy density
at some time prior to its decay,
which marks the beginning of yet another stage of matter domination.
The decay of the axion into relativistic gauge bosons and the
corresponding renewed transition to radiation domination
then represent a second installment of reheating, which can 
be described by the same set of equations as the
primary reheating process, cf.\ Eqs.~\eqref{eq:phiradBoltz}
and \eqref{eq:Friedmann}.
With the axion decay rate 
\begin{align}
\Gamma_a \simeq \frac{\alpha^2}{64\pi^3} \frac{m_a^3}{f_a^2} \,, \quad
\alpha = \frac{g_2^2}{4\pi} \,,
\end{align}
and using Eq.~\eqref{eq:TRH}, we find for the
{\em secondary reheating temperature} or axion decay temperature 
\begin{align}
T_{\rm dec} \simeq 1 \times 10^{4}\,\text{GeV}
\bigg(\frac{m_a}{10^9\,\text{GeV}}\bigg)^{3/2}
\left(\frac{10^{15}\,\text{GeV}}{f_a}\right)\,.
\end{align}
This temperature should be at least of $\mathcal{O}(10)\,\text{MeV}$~\cite{low_reheat},
which imposes a lower bound on $m_a$: 
\begin{align}
m_a \gtrsim 8 \times 10^4\,\textrm{GeV} \left(\frac{f_a}{10^{15}\,\text{GeV}}\right)^{2/3} \,.
\label{eq:mabound2}
\end{align}

%%%%%%%%%%%%%%%%%%%%%%%%%%%%%%%%%%%%%%%%%%%%%%%%%%%%%%%%%%%%%%%%%%%%%%%%%%%%%%%%%%%%%%%%%%%%%%%%%%%%

The five differential equations in
Eqs.~\eqref{eq:axionEOM}, \eqref{eq:LBoltz}, \eqref{eq:phiradBoltz}, and
\eqref{eq:Friedmann}  allow one to compute the present value of the 
baryon asymmetry (i.e., the baryon-to-photon ratio),
\begin{align}
\eta_B^0 \equiv \frac{n_B^0}{n_\gamma^0} =
c_{\rm sph} \, \frac{g_{*,s}^0}{g_*} \,\eta_L^a  \simeq 0.013 \,\eta_L^a\,,
\label{eq:etaB}
\end{align}
where the sphaleron factor $c_{\rm sph}$ accounts for
the conversion of the lepton asymmetry into baryon asymmetry by sphalerons.
Here, $g_{*,s}^0$ and $g_*$ denote the effective numbers
of degrees of freedom contributing to the entropy density 
in the present epoch and during reheating, respectively.
In the standard model 
$c_{\rm sph} = 28/79$, $g_{*,s}^0 = 43/11$, and $g_* = 427/4$.
Last but not least, $\eta_L^a$ in Eq.~\eqref{eq:etaB} stands for
the final lepton asymmetry after the decay of the axion
around $t \simeq \Gamma_a^{-1}$.

%%%%%%%%%%%%%%%%%%%%%%%%%%%%%%%%%%%%%%%%%%%%%%%%%%%%%%%%%%%%%%%%%%%%%%%%%%%%%%%%%%%%%%%%%%%%%%%%%%%%

\begin{figure}
\includegraphics[width=0.45\textwidth]{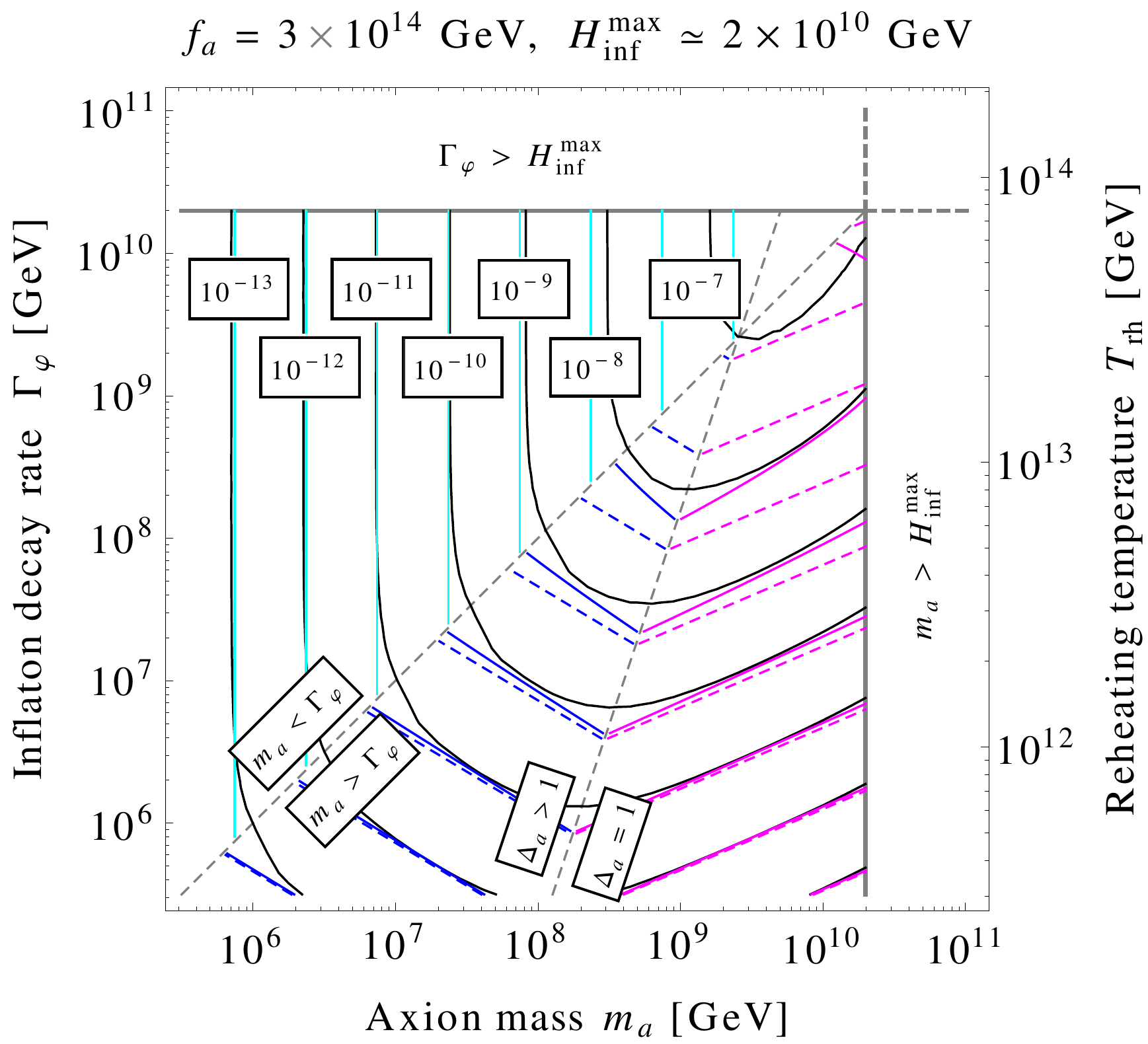}
\caption{Contour plot of the final baryon asymmetry $\eta_B^0$
as a function of the axion mass $m_a$ and the inflaton decay
rate $\Gamma_\varphi$ for an axion decay constant of $f_a = 3 \times 10^{14}\,\text{GeV}$.
The black (bent) contours represent the full numerical result, while
the colorful (straight) contours depict our analytical estimate
according to Eqs.~\eqref{eq:etaB} and \eqref{eq:etaLa}.
In the lower part of the plot ($m_a > \Gamma_\varphi$), the effect of washout
is illustrated by the difference between the dashed ($\kappa = 0$) and solid
($\kappa \neq 0$) lines.
}
\label{fig:etacontours}
\end{figure}

%%%%%%%%%%%%%%%%%%%%%%%%%%%%%%%%%%%%%%%%%%%%%%%%%%%%%%%%%%%%%%%%%%%%%%%%%%%%%%%%%%%%%%%%%%%%%%%%%%%%

We determine $\eta_L^a$ by solving the five differential equations in
Eqs.~\eqref{eq:axionEOM}, \eqref{eq:LBoltz}, \eqref{eq:phiradBoltz}, and
\eqref{eq:Friedmann} numerically.
We also present approximate analytical solutions,
which will be discussed in detail in an upcoming publication. 
It is convenient to parametrize $\eta_L^a$ as follows: 
\begin{align}
\eta_L^a = C \, \Delta_a^{-1} \Delta_\varphi^{-1}\,\eta_L^{\rm max}\, e^{-\kappa} \,.
\label{eq:etaLa}
\end{align}
The approximate analytical results agree with the numerical results,
as shown in Fig.~\ref{fig:etacontours}. 
In the following, we shall present analytical expressions for the
individual factors on the right-hand side of Eq.~\eqref{eq:etaLa}.
$\eta_L^{\rm max}$ denotes the all-time maximum value of the lepton 
asymmetry, which is reached around the time when the axion oscillations set it,
i.e., at $t \sim t_{\rm osc} \simeq m_a^{-1}$.
Note that this time does not necessarily coincide with the time when $\Gamma_L \simeq H$.
Integrating the Boltzmann equation for the lepton asymmetry
up to $t \sim t_{\rm osc}$, one approximately finds 
\begin{align}
\eta_L^{\rm max} \simeq \frac{\sigma_{\rm eff}}{g_*^{1/2}} \frac{a_0}{f_a}
\, m_a \, M_{\rm Pl}
\times \min\left\{1,\,\left(\Gamma_\varphi/m_a\right)^{1/2}\right\}
\,, \label{eq:etaLmax}
\end{align}
which is suppressed w.r.t.\ the would-be equilibrium
lepton asymmetry, $\eta_L^{\rm eq} = n_L^{\rm eq}/n_\gamma^{\rm eq}$,
evaluated at the same time by a factor $n_L / n_L^{\rm eq} \simeq
T / T_L \times \min\big\{1,\,\left(\Gamma_\varphi/m_a\right)^{1/2}\big\}$.
Remarkably enough, the maximal lepton asymmetry is rather insensitive
to the axion decay constant; it only depends on the 
ratio $a_0/f_a$, which is expected to be $\mathcal{O}(1)$.
Furthermore, $\eta_L^{\rm max}$ turns out to be directly proportional to
the effective cross section $\sigma_{\rm eff}$.
For $a_0 = f_a$ and given the value of $\sigma_{\rm eff}$ in Eq.~\eqref{eq:sigmaeff},
$\eta_L^{\rm max}$ is, hence, typically much larger than the observed
value, $\eta_B^{\rm obs} \simeq 6\times10^{-10}$~\cite{Ade:2013zuv},
\begin{align}
\eta_L^{\rm max} \simeq 2 \times 10^{-5}
\bigg(\frac{m_a}{10^9\,\text{GeV}}\bigg)^{p}
\left(\frac{\Gamma_\varphi}{10^9\,\text{GeV}}\right)^{q} \,,
\end{align}
where the powers $p$ and $q$ are given as in Eq.~\eqref{eq:etaLmax}.

%%%%%%%%%%%%%%%%%%%%%%%%%%%%%%%%%%%%%%%%%%%%%%%%%%%%%%%%%%%%%%%%%%%%%%%%%%%%%%%%%%%%%%%%%%%%%%%%%%%%

\begin{figure}
\includegraphics[width=0.45\textwidth]{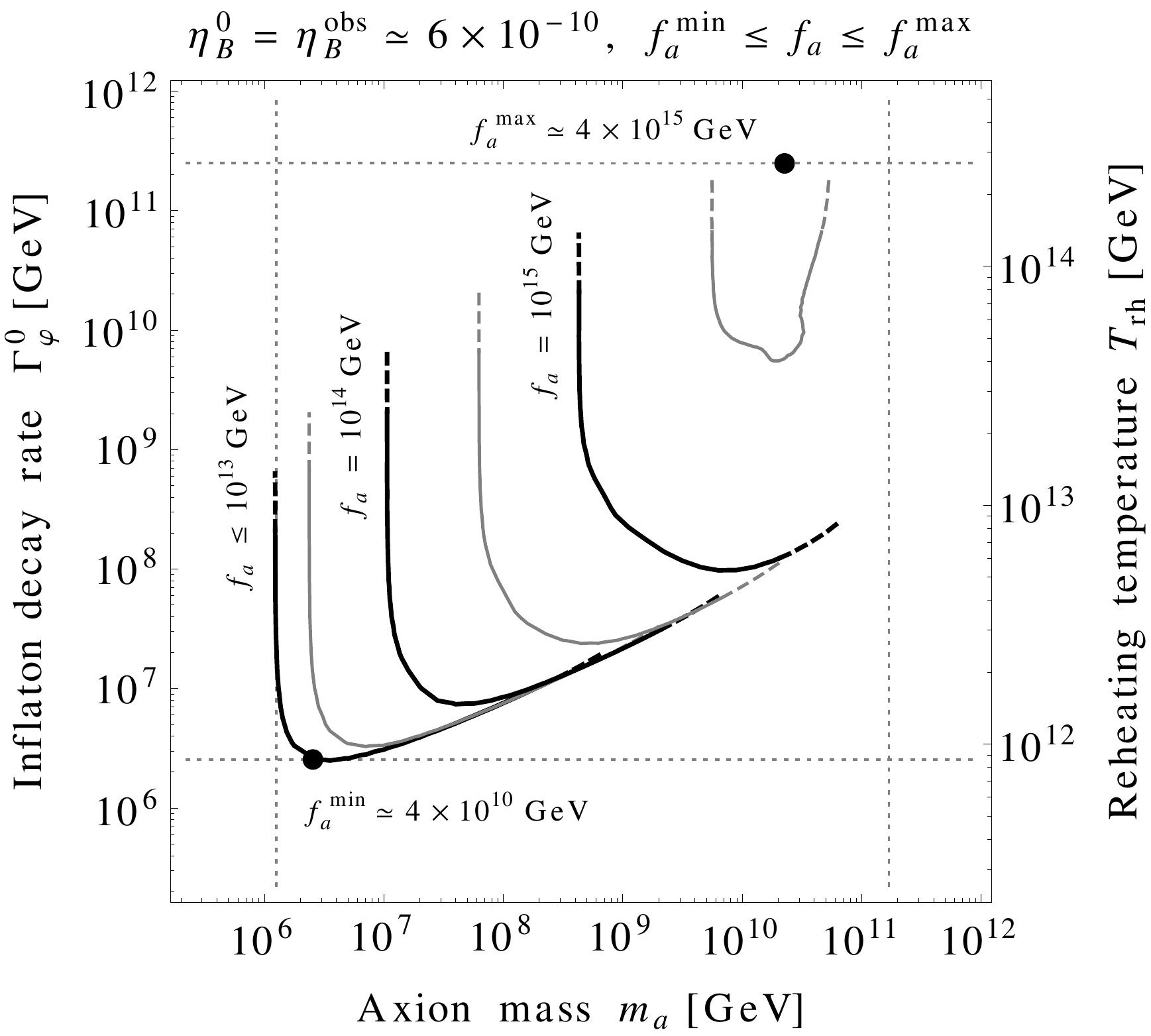}
\caption{Contour lines for successful leptogenesis ($\eta_B^0 = \eta_B^{\rm obs}$)
in the $m_a$--$\Gamma_\varphi$ plane for different values of the axion decay constant $f_a$.
The dashed segments along the individual contours mark the regions where either $m_a$
or $\Gamma_\varphi$ become comparable to the maximally allowed Hubble rate
$H_{\rm inf}^{\rm max}$, cf.\ Eq.~\eqref{eq:Hbound}.
For $f_a \lesssim 10^{13}\,\textrm{GeV}$, entropy production in axion decays ceases to affect
$\eta_B^0$, cf.\ Eq.~\eqref{eq:Deltaa}, which is reflected in the contour lines being no longer
sensitive to changes in $f_a$.
The lower bounds on $m_a$ and $f_a$ then directly follow from our constraints in
Eqs.~\eqref{eq:Hbound}, \eqref{eq:mabound}, and \eqref{eq:Gammabound}.
At the same time, the regime of large $f_a$, and hence the upper
bounds on $m_a$ and $f_a$, require a careful numerical analysis due to the
strong impact of washout.}
\label{fig:facontours}
\end{figure}

The two $\Delta$ factors in Eq.~\eqref{eq:etaLa} account for the entropy
production in inflaton and axion decays during reheating and at late
times, respectively.
We approximately have
\begin{align}
\Delta_\varphi \simeq \max\left\{1,\,\Delta_\varphi'\right\} \,, \quad
\Delta_a \simeq \max\left\{1,\,\Delta_a'\right\} \,,
\end{align}
where $\Delta_\varphi' \simeq \left(m_a/\Gamma_\varphi\right)^{5/4}$ and
with $\Delta_a'$ being given as
\begin{align}
\Delta_a' \simeq 
\frac{2\pi^2}{\alpha} \frac{f_a\,a_0^2}{m_a M_{\rm Pl}^2} 
\times \min\left\{1,\,\left(\Gamma_\varphi/m_a\right)^{1/2}\right\}
\,. \label{eq:Deltaa}
\end{align}
In the region of parameter space in which we are able to
successfully reproduce $\eta_B^{\rm obs}$, 
entropy production in axion decays begins to play a
role for $f_a$ values around $3\times 10^{13}\,\textrm{GeV}$,
cf.\ Fig.~\ref{fig:facontours}.
For smaller values of $f_a$, we always have $\Delta_a = 1$
in the entire parameter region of interest.

%%%%%%%%%%%%%%%%%%%%%%%%%%%%%%%%%%%%%%%%%%%%%%%%%%%%%%%%%%%%%%%%%%%%%%%%%%%%%%%%%%%%%%%%%%%%%%%%%%%%

%%%%%%%%%%%%%%%%%%%%%%%%%%%%%%%%%%%%%%%%%%%%%%%%%%%%%%%%%%%%%%%%%%%%%%%%%%%%%%%%%%%%%%%%%%%%%%%%%%%%

The factor $e^{-\kappa}$ in Eq.~\eqref{eq:etaLa}
accounts for the washout of $\eta_L^{\rm max}$ during
reheating due to the  $\Delta L = 2$ washout processes,
cf.\ the term proportional to $-\sigma_{\rm eff}\, n_L$ in Eq.~\eqref{eq:LBoltz}.
In case the axion begins to oscillate before the end of reheating,
i.e., for $m_a \gtrsim \Gamma_\varphi$, one can estimate
\begin{align}
\kappa \sim \frac{T_{\rm rh}}{T_L} \simeq 1 \left(\frac{T_{\rm rh}}{10^{13}\,\text{GeV}}\right) \,.
\end{align}
For $m_a \lesssim \Gamma_\varphi$ on the other hand, washout is always
negligible, so that $\kappa$ can be safely set to $\kappa = 0$.
A more careful treatment of the effect of washout on the final baryon asymmetry
in our scenario is left for future work.

%%%%%%%%%%%%%%%%%%%%%%%%%%%%%%%%%%%%%%%%%%%%%%%%%%%%%%%%%%%%%%%%%%%%%%%%%%%%%%%%%%%%%%%%%%%%%%%%%%%%

Finally, the factor $C$ in Eq.~\eqref{eq:etaLa} is a numerical fudge factor, which can,
in principle, be estimated analytically, but
which, in practice, is best determined by fitting $\eta_L^a$ in
Eq.~\eqref{eq:etaLa} to the outcome of our numerical analysis.
Specifically, we find $C\simeq 1.5$ for $m_a \lesssim \Gamma_\varphi$ and
$C\simeq 2.2$ for $m_a \gtrsim \Gamma_\varphi$.
The fact that these values are both of $\mathcal{O}(1)$ confirms
the accuracy of our analytical estimate.

Altogether, the parameter dependence of the final lepton asymmetry in
Eq.~\eqref{eq:etaLa} can be summarized as follows (here, we neglect the effect
of washout and set $\kappa\rightarrow 0$),
\begin{align}\label{eq:final}
\eta_L^a \propto
\frac{1}{T_L}
\begin{cases}
m_a^{-3/4} \, \Gamma_\varphi^{7/4} \, a_0 \, f_a^{-1} & \hspace{-8pt}; \:
m_a \gtrsim \Gamma_\varphi ,\, \Delta_a = 1 \\
m_a \, a_0 \, f_a^{-1} & \hspace{-8pt}; \:
m_a \lesssim \Gamma_\varphi ,\, \Delta_a = 1 \\
m_a^{3/4} \, \Gamma_\varphi^{5/4} \, M_{\rm Pl}^2 \, a_0^{-1} \, f_a^{-2} & \hspace{-8pt}; \:
m_a \gtrsim \Gamma_\varphi ,\, \Delta_a > 1 \\
m_a^2 \, M_{\rm Pl}^2 \, a_0^{-1} \, f_a^{-2} & \hspace{-8pt}; \:
m_a \lesssim \Gamma_\varphi ,\, \Delta_a > 1
\end{cases} \,.
\end{align}
In the various regions of parameter space, typical values for $m_a$,
$\Gamma_\varphi$, and $f_a$ then yield the following
final baryon asymmetries
(again for $a_0 = f_a$ and $\sigma_{\rm eff}$ as in Eq.~\eqref{eq:sigmaeff}),

%%%%%%%%%%%%%%%%%%%%%%%%%%%%%%%%%%%%%%%%%%%%%%%%%%%%%%%%%%%%%%%%%%%%%%%%%%%%%%%%%%%%%%%%%%%%%%%%%%%

\begin{center}\begin{tabular}{ccccc}
$f_a$ [GeV] & $m_a$ [GeV] &  $\Gamma_\varphi$ [GeV] & $\eta_B^0$ & $\Delta_a$
\\\hline\hline\noalign{\vskip 1mm}  
$10^{12}$ & $3\times10^{7\phantom{0}}$ & $3\times10^{6\phantom{0}}$
          & $3 \times 10^{-10}$ & $1$ \\
$10^{12}$ & $3\times10^{6\phantom{0}}$ & $3\times10^{7\phantom{0}}$ 
          & $1 \times 10^{-9\phantom{0}}$ & $1$ \\
$10^{15}$ & $1\times10^{10}$ & $1\times10^{9\phantom{0}}$ 
          & $7 \times 10^{-9\phantom{0}}$ & $3$ \\
$10^{15}$ &  $1\times10^{9\phantom{0}}$ & $1\times10^{10}$
          & $6 \times 10^{-9\phantom{0}}$ & $80$
\end{tabular}
\end{center}

%%%%%%%%%%%%%%%%%%%%%%%%%%%%%%%%%%%%%%%%%%%%%%%%%%%%%%%%%%%%%%%%%%%%%%%%%%%%%%%%%%%%%%%%%%%%%%%%%%%%

Let us now determine the range of parameters that admit the
correct value of the baryon asymmetry in view of 
the constraints in Eqs.~\eqref{eq:Hbound}, \eqref{eq:mabound},
\eqref{eq:Gammabound}, and \eqref{eq:mabound2}, cf.\
Fig.~\ref{fig:facontours}, which shows the contour lines
of successful leptogenesis for different values of $f_a$.
The range of allowed values spans five orders of magnitude,
\begin{align}
4\times 10^{10}\,\text{GeV} \lesssim f_a \lesssim 4 \times 10^{15}\,\text{GeV}.
\label{eq:farange}
\end{align}
For smaller values of $f_a$, it is not possible to
generate a sufficiently large baron asymmetry, while
keeping the baryonic isocurvature perturbations small enough.
For larger values of $f_a$, the dilution of the asymmetry during the
late-time decay of the axion is too strong.
Varying $f_a$ within the interval in Eq.~\eqref{eq:farange},
we then find that $m_a$, $\Gamma_\varphi$, and $T_{\rm rh}$
can take values within the following ranges: 
\begin{align}
1 \times 10^6\,\text{GeV} \lesssim & \:\, m_a \lesssim 2 \times 10^{11}\,\text{GeV} \,, \\
3 \times 10^6\,\text{GeV} \lesssim & \:\,\Gamma_\varphi \, \lesssim 3 \times 10^{11}\,\text{GeV} \,, \\
9 \times 10^{11}\,\text{GeV} \lesssim & \:\, T_{\rm rh} \lesssim 3 \times 10^{14}\,\text{GeV} \,.
\end{align}
These ranges of parameters are consistent with models of
dynamical axions, as well as string axion models.

%%%%%%%%%%%%%%%%%%%%%%%%%%%%%%%%%%%%%%%%%%%%%%%%%%%%%%%%%%%%%%%%%%%%%%%%%%%%%%%%%%%%%%%%%%%%%%%%%%%%

Finally, let us conclude.
As we have been able to show, lepton number violation due to the exchange
of heavy Majorana neutrinos, in combination with the effective chemical potential 
generated by a slowly rolling axionlike scalar field, is sufficient for a successful
realization of baryogenesis via leptogenesis.
In this scenario, the baryon asymmetry does neither depend on the concrete
heavy neutrino mass spectrum nor on the amount of $CP$ violation
in the light and heavy neutrino sectors.
In particular, it is consistent with (almost) degenerate heavy
neutrino masses close to the GUT scale.
Hence, while thermal leptogenesis assumes some of the heavy neutrino Yukawa couplings
to be much smaller than $\mathcal{O}(1)$, our mechanism
equally applies in the case of Yukawa couplings of $\mathcal{O}\left(10^{-1}\cdots 1\right)$.
Furthermore, as can be seen from Eqs.~\eqref{eq:sigmaeff} and \eqref{eq:final}, 
the baryon asymmetry increases with the light neutrino masses $m_i$.
Thus, while thermal leptogenesis imposes an upper bound on the neutrino mass scale,
$\bar{m} \lesssim 0.2\,\text{eV}$, to avoid too strong washout~\cite{Buchmuller:2005eh},
our scenario works for all experimentally allowed light neutrino masses.
This bound will soon be probed by a multitude of terrestrial experiments~\cite{Drexlin:2013lha}
as well as in cosmological and astrophysical observations~\cite{Abazajian:2011dt}.
The axion-driven leptogenesis mechanism presented in this Letter therefore appears
to be an attractive alternative to the conventional scenario of thermal
leptogenesis.
On the other hand, further work is clearly needed to embed it into a more complete model.
Here, we expect our mechanism to yield a number of nontrivial
model-building constraints due to some of its peculiar features:
(i) the requirement of a high reheating temperature, (ii) the fact that $f_a$
may potentially lie in the vicinity of the decay constant of the QCD axion,
(iii) the possibility of baryonic isocurvature perturbations at a detectable
level, etc.
Such constraints may easily allow for a possibility to test our mechanism
in the near future and assess whether it has indeed the potential to serve
as a viable explanation for the baryon asymmetry of the Universe.

%%%%%%%%%%%%%%%%%%%%%%%%%%%%%%%%%%%%%%%%%%%%%%%%%%%%%%%%%%%%%%%%%%%%%%%%%%%%%%%%%%%%%%%%%%%%%%%%%%%%

The authors wish to thank K.~Harigaya, A.~Kamada, K.~Kamada, M.~Kawasaki, L.~Pearce,
and F.~Takahashi for helpful comments and discussion.
This work has been supported in part by the U.\,S.\ Department of Energy Grant DE-SC0009937 (A.\,K.),
by Grants-in-Aid for Scientific Research from the Ministry of Education, Culture, Sports,
Science, and Technology (MEXT), Japan, No.\ 26104009 and No.\ 26287039 (T.\,T.\,Y.), and by
the World Premier International Research Center Initiative (WPI), MEXT, Japan.

%%%%%%%%%%%%%%%%%%%%%%%%%%%%%%%%%%%%%%%%%%%%%%%%%%%%%%%%%%%%%%%%%%%%%%%%%%%%%%%%%%%%%%%%%%%%%%%%%%%%

%%%%%%%%%%%%%%%%%%%%%%%%%%%%%%%%%%%%%%%%%%%%%%%%%%%%%%%%%%%%%%%%%%%%%%%%%%%%%%%%%%%%%%%%%%%%%%%%%%%%

\end{document}